\documentclass[%
 reprint,
 amsmath,amssymb,
 aps,
superscriptaddress,
prb,nofootinbib
]{revtex4-2}
\usepackage{tikz}
\usepackage{float}
\usepackage{siunitx}
\usepackage{hyperref}
\usepackage{physics}
\usepackage[capitalise]{cleveref}
\usepackage{comment}
\usepackage{nicefrac}
\usepackage[percent]{overpic}
\usepackage{array, makecell}

\usepackage{amsmath}
\usepackage{graphicx}
\usepackage{dcolumn}
\usepackage{bm}
\usepackage{subfiles}
\usepackage{floatflt}

\begin{document}
\title{Micromagnetic study of spin transport in easy-plane antiferromagnetic insulators}

\author{Verena Brehm}
 \email{verena.j.brehm@ntnu.no}
\affiliation{ Center for Quantum Spintronics, Department of Physics, Norwegian University of Science and Technology, 7491 Trondheim, Norway}%
\author{Olena Gomonay}%
\affiliation{Institute of Physics, Johannes Gutenberg-University Mainz, Staudingerweg 7, Mainz 55128, Germany}%
\author{Serban Lepadatu}%
\affiliation{Jeremiah Horrocks Institute for Mathematics, Physics and Astronomy, University of Central Lancashire, Preston, PR1 2HE, United Kingdom}%
\author{Mathias Kl\"{a}ui}%
\affiliation{ Center for Quantum Spintronics, Department of Physics, Norwegian University of Science and Technology, Trondheim, Norway}%
\affiliation{Institute of Physics, Johannes Gutenberg-University Mainz, Staudingerweg 7, Mainz 55128, Germany}%
\author{Jairo Sinova}%
\affiliation{Institute of Physics, Johannes Gutenberg-University Mainz, Staudingerweg 7, Mainz 55128, Germany}%
\author{Arne Brataas}%
\affiliation{ Center for Quantum Spintronics, Department of Physics, Norwegian University of Science and Technology, Trondheim, Norway}%
\author{Alireza Qaiumzadeh}%
\affiliation{ Center for Quantum Spintronics, Department of Physics, Norwegian University of Science and Technology, Trondheim, Norway}%

\date{\today}

\begin{abstract}
Magnon eigenmodes in easy-plane antiferromagnetic insulators are linearly polarized and are not expected to carry any net spin angular momentum. Motivated by recent nonlocal spin transport experiments in the easy-plane phase of hematite, we perform a series of micromagnetic simulations in a nonlocal geometry at finite temperatures. 
We show that by tuning an external magnetic field, we can control the magnon eigenmodes and the polarization of the spin transport signal in these systems.  
We argue that a \textit{coherent beating oscillation} between two orthogonal linearly polarized magnon eigenmodes is the mechanism responsible for finite spin transport in easy-plane antiferromagnetic insulators. The sign of the detected spin signal  is also naturally explained by the proposed coherent beating mechanism. Our finding opens a path for on-demand control of the spin signal in a large class of easy-plane antiferromagnetic insulators. 

\end{abstract}

\maketitle

\section{Introduction} 
    On-demand control and long-distance transport of spin angular momentum in antiferromagnetic insulators (AFMIs) is among the cornerstones of modern spintronics. Negligible stray fields, operating at THz frequencies, and the lack of Joule heating in AFMI make them suitable candidates for the miniaturization of next-generation ultrafast spintronic-logic devices \cite{AFMreviewManchon,AFMreviewHelen}.

    In collinear AFMIs with uniaxial easy-axis magnetic anisotropy, such as hematite below the Morin transition temperature \cite{MorrishHematite}, two circularly polarized magnon eigenmodes have opposite helicity. Each circularly polarized magnon mode carries one unit of spin angular momentum $\pm \hbar$. This spin angular momentum can be transported across micrometer distances, as recently demonstrated in nonlocal spin detection experiments in the \textit{easy-axis} collinear phase of hematite $\alpha$-Fe$_2$O$_3$  \cite{tunableLongDistanceKlaui,2020NanoL..20..306R}. The sign of the spin accumulation in nonlocal spin transport measurements encodes the  polarization of the transmitted spin angular momentum via magnons. 

    On the contrary, the two orthogonal magnon eigenmodes in the \textit{easy-plane} AFMIs are linearly polarized and thus typically do not carry any net spin angular momentum. However, recently, Refs. \onlinecite{lebrun_long-distance_2020,han_birefringence-like_2020,2020ApPhL.117x2405R,2022JMMM..54368631R,akash1} reported the transport of spin angular momentum on the micrometer scale in the easy-plane phase of hematite. Above the Morin transition temperature, hematite is a canted AFMI or weak ferromagnet with easy-plane anisotropy  \cite{MorrishHematite}. 
    Some of these experiments also showed a  magnetic field and spatial-dependent sign change of the spin accumulation, and thus of the magnon polarization \cite{2020ApPhL.117x2405R,akash1}.  

    In Refs \cite{lebrun_long-distance_2020,han_birefringence-like_2020}, the authors attributed the detected finite spin angular momentum transport to a birefringence--like mechanism. Within this framework, the origin of the sign change of the spin signal remained unknown. On the other hand, in Refs. \cite{akash1,akash2} the authors attributed both finite spin angular momentum transport and the change in the sign of the spin signal to a Hanle-like behavior of the magnon pseudospin. However, within their formalism, the coupling between a small magnetization, induced by a homogeneous  Dzyaloshinskii-Moriya (DM) interaction, and applied transverse magnetic field plays a crucial role in explaining the sign change of the spin signal \cite{akash1,AkashMajorTheory}.
 
    In this paper, we perform finite-temperature micromagnetic simulations to study the spin angular momentum transport mechanism as well as the distance- and magnetic-field-dependent sign change of the spin signal in easy-plane AFMI systems. We explain our numerical observations with a coherent beating mechanism between two linearly polarized AFM magnon eigenmodes. We argue that both the finite spin transport and the magnetic field- and spatial-dependent sign of the spin signal are generic features of all easy-plane AFMIs and not only weak ferromagnets or canted AFMIs with a finite DM interaction, such as easy-plane hematite.

   This paper is organized as follows. After introducing our model and simulation technique in \cref{sec:theo}, we will first demonstrate the modulation of the magnon dispersion relation through an external magnetic field in \cref{sec:disprel}. Second, in \cref{sec:beating} we will propose our coherent beating oscillation mechanism that is based on the pairing of magnons on the two dispersion branches, and demonstrate signatures of these pairs numerically in \cref{sec:numbeating}. Using this established model, we explain the numerically obtained spin transport signal and its oscillatory nature in \cref{sec:transport}. Finally, we conclude our findings in \cref{sec:conclusion}.
    
\section{Model} \label{sec:theo}
   
    \begin{figure}
        \centering
        \includegraphics[width=1.0\linewidth]{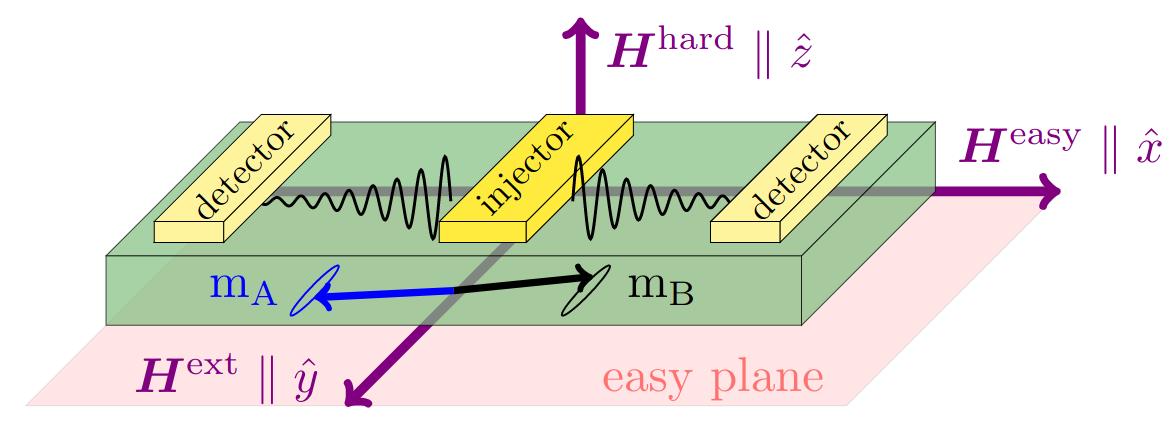}
        \caption{Schematic presentation of the setup. Spin current is injected into an easy-plane AFMI (green region) by the central electrode via spin Hall torque and measured by the electrodes on the  left and right via the inverse spin Hall effect. Purple arrows show the orientation of easy $\bm{H}^\mathrm{easy}$, and hard $\bm{H}^\mathrm{hard}$, magnetic anisotropy fields. $\bm{H}^\mathrm{ext}$ is a transverse external magnetic field, used to modulate the spin transport signal. The black and blue arrows show the two AFM sublattices with elliptical precession.}
        \label{fig:OurSetup}
    \end{figure}
   
    \subsection{System setup}
    \label{sec:model}
    We model a thin layer of AFMI in an orthorhombic phase using two anisotropy axes (see \cref{fig:OurSetup}). The hard-axis anisotropy  is set along the $z$ direction, which gives an easy-plane magnetic anisotropy in the sample plane ($x$-$y$ plane). A much weaker easy-axis magnetic anisotropy lies within the easy plane and gives a defined ground state along the $x$ direction. In addition, a transverse magnetic field, $\bm{H}^\text{ext}=H_y \hat{y}$, perpendicular to both the easy and hard magnetic anisotropy axis, is applied to modulate the AFM magnon dispersion, as shown later. Besides, we add a homogeneous DM interaction parallel to the hard axis, and investigate its effect on the magnon dispersion and transport.

    In order to simulate spin transport, magnons are pumped in the center via a spin Hall spin-transfer torque and propagate along the $x$ direction to the left and right. The spin signal is recorded with detectors that are placed at various distances along the $x$ direction. The polarization-  and distance-dependent spin transport signal is recorded for each strength of the transverse magnetic field $H_y$.

    \begin{table}[htbp]
	\centering
	\caption{Micromagnetic parameters of hematite, used for  simulations.}
	\begin{tabular}{@{}llll@{}} \hline
		\thead{\textbf{Quantity}}                 & \thead{\textbf{Symbol}}            & \thead{\textbf{Value}}                & \thead{\textbf{Unit}}                 \\ 
		Length of AFMI layer	& $L_x$ & $\SI{2.5}{}$ & $\si{\micro \meter}$ \\
		Width of AFMI layer	& $L_y$ & $\SI{0.1}{}$ & $\si{\micro \meter}$ \\
		Thickness of AFMI layer	& $L_z$ & $\SI{5}{}$ & $\si{\nano \meter} $\\
		Grid size	& a & $\SI{5}{}$ & $\si{\nano \meter} $\\
		Exchange stiffness	& $A_\text{AFM}$                      &  $\SI{76}{}$                    &         $\si{\femto\joule\per\meter}$ \\
		Homogeneous exchange	& $a_\text{AFM}$                      &  $\SI{-460}{}$                    &         $\si{\kilo\joule \per \meter\cubed}$ \\
		Easy-axis anisotropy & $K_\text{easy}$                      &    $-\SI{21}{}$                  &            $\si{\milli\joule \per \meter\cubed}$         \\
		Hard-axis anisotropy & $K_\text{hard}$ 	 & $\SI{21}{}$ & $\si{\joule \per \meter\cubed}$ \\
		Saturation magnetization & $M_s$                      &      $\SI{2.1}{}$                  &                 $\si{\kilo\ampere\per \meter}$ \\ 
		Gilbert damping&           $\alpha$           &   $\SI{2e-4}{}$                    &                      1\\ 
		Homogeneous DM interaction &           $D$            &        $\SI{4.6}{}$             &                    $\si{\kilo\joule \per \meter\cubed}$  \\
		Time step &           $\Delta t$            &      \thead{$\SI{1}{}$ @ $T=0$ \\ $\SI{0.5}{}$ @ $T \neq 0$}             &                    $\si{\femto\second}$  \\
        Charge current density for SOT & $J_c$ & \SI{560}{} & \si{\mega\ampere\per\meter\squared} 
	\end{tabular}
	\label{tab:coefficients}
    \end{table}

    Without loss of generality, we use easy-plane hematite parameters in our micromagnetic simulations. The simulation parameters \cite{sulymenko_terahertz-frequency_2017} are listed in \cref{tab:coefficients}. The characteristic length scales, at zero magnetic field, given by the exchange stiffness and the anisotropy constants $\lambda_l=\sqrt{A_\text{AFM}/(2K_l)}$, are $\lambda_\text{hard} \approx \SI{43}{\nano\meter}$ and $\lambda_\text{easy}\approx \SI{1.35}{\micro\meter}$ for the hard- and easy-axis anisotropies, respectively. In order to avoid the reflection of magnons at the edges of the system, we use a Gilbert damping parameter and system length that ensure the excitation decays to zero before. 

    \subsection{Micromagnetic simulations}
    Finite temperature simulations are conducted using a stochastic micromagnetic framework implemented in the open-source code \textsc{BORIS} \cite{BORIS}. Within the micromagnetic assumption, every discrete simulation cell with a volume $V$ is assigned a macrospin magnetic moment $\bm{M}$ with a homogeneous saturation magnetization $M_s$ by averaging over all magnetic contributions \cite{AtomMicroScales}. Under the influence of temperature $T$, the dynamics of the magnetic moment direction $\bm{m} = \bm{M}/M_s$, in a two-sublattice AFMI, is described by coupled stochastic Landau-Lifshitz-Gilbert (sLLG) equations,
    \begin{align}
        d_t \bm{m}_i &= - \gamma \bm{m}_i \times \left(\bm{H}_i^\text{eff} + \bm{H}^\text{th}_i\right) \nonumber \\&- \alpha \gamma \bm{m}_i \times \left[ \bm{m}_i \times \left(\bm{H}_i^\text{eff} + \bm{H}^\text{th}_i\right)\right] ,
    \end{align} 
    where $i \in \{A,B\}$ refers to two AFM sublattices. We use $\gamma = \mu_0 \vert \gamma_e \vert / (1+\alpha^2)$ with the vacuum permeability $\mu_0$, the electron gyromagnetic ratio $\gamma_e = - {g\mu_B}/{\hbar}$ with the electron $g$-factor $g$ and the Bohr magneton $\mu_B$, the reduced Planck constant $\hbar$, and the dimensionless Gilbert damping parameter $\alpha$ \cite{ASDeriksson}.

    $\bm{H}^\text{eff}$ denotes the effective magnetic field at the magnetic site $i$, and $\bm{H}^\text{th}$ is a stochastic thermal field that adds temperature to the model. For the latter, a normalized Gaussian distribution is scaled with the prefactor
    $\xi_{th}= \sqrt{{2\alpha k_B T}/{(\gamma \mu_0 M_s V \Delta t)}}$ in every component adding white noise to the system that is weighted with the thermal energy $k_B T$, with $k_B$ the Boltzmann constant, and scaled with both the cell size volume $V$ and the time step of the simulation  $\Delta t$. 
    
    In our model, the effective magnetic field is given by
    \begin{equation}
    \bm{H}_i^\text{eff} = \bm{H}_i^\text{ex} + \bm{H}_i^\text{DM} + \bm{H}_i^\text{aniso} + \bm{H}_i^\text{ext} + \bm{H}_i^\text{SOT}. 
    \end{equation}
    $\bm{H}_i^\text{ex}$ is the sum of homogeneous and inhomogeneous exchange interactions \cite{BORIS},
    \begin{align}
        \bm{H}_{i}^\text{ex} = - \frac{4 a_\text{AFM}}{\mu_0 M_s} \left[\bm{m}_i \times \left(\bm{m}_i \times \bm{m}_j \right)\right]+\frac{2 A_\text{AFM}}{\mu_0 M_s} \nabla^2 \bm{m}_i 
    \end{align} 
    where $i \neq j$, $a_\text{AFM}$ is the homogeneous exchange constant, and $A_\text{AFM}$ is the AFM exchange stiffness.
    
    $\bm{H}_i^\text{DM}$ is the homogeneous DM interaction field \cite{BORIS},
    \begin{align}
        \bm{H}_{i}^\text{DM} =  - \eta_i \frac{D}{\mu_0 M_s} \bm{d}_h \times \bm{m}_j,
    \end{align} 
    where $\eta_{A(B)}=+(-)1$, $D$ is the homogeneous DM interaction strength, and $\bm{d}_h$ is the direction of the DM vector.
    
    $\bm{H}_i^\text{aniso}$ is the anisotropy field \cite{BORIS},
    \begin{align}
        \bm{H}_i^\text{aniso} = \bm{H}_i^\text{hard} + \bm{H}_i^\text{easy} = \sum_{l \in \{\text{hard},\text{easy}\}} \frac{2 K_l}{\mu_0 M_s} \left(\bm{m}_i \cdot \hat{e}_l\right) \hat{e}_l,
    \end{align} 
    where the hard-axis anisotropy is given by $K_\text{hard} > 0$ and $\hat{e}_\text{hard} = \hat{z}$, and the easy-axis anisotropy is along $\hat{e}_\text{easy}=\hat{x}$ with $K_\text{easy} < 0$. 
    
    $\bm{H}_i^\text{ext}$ is the external magnetic field that couples to AFM spins via a Zeeman coupling mechanism. In our simulations, we apply a dc magnetic field perpendicular to both the easy-axis and the hard-axis anisotropy fields, i.e., along the $y$ direction. Therefore we call it the \textit{transverse magnetic field $H_y$}.
    
    Finally, $\bm{H}_i^\text{SOT}$ is the total spin-orbit torque (SOT), which is the sum of a fieldlike and a dampinglike torque \cite{BORIS},
    \begin{align} \label{eq:torque}
        \bm{H}_i^\text{SOT} = - \frac{\Theta}{\gamma M_s}  \frac{\mu_B}{e}\frac{|J_c|}{L_z}\left(\bm{m}_i \times \bm{P} + r_G \bm{P} \right),
    \end{align} 
    generated by a charge current with the density $J_c$ that is converted to a spin current via the spin Hall effect. $\bm{P}$ is the direction of spin-Hall-induced spin polarization at the interface. Furthermore, $\Theta$ is the spin Hall angle, a measure of the efficiency of the spin-to-charge current conversion, and $r_G$ parameterizes the fieldlike torque amplitude. 
    
    In order to model the spin-Hall-induced SOT (see \cref{fig:OurSetup}), we set $\Theta> 0$ in the injector region, that lies in the center of the system and has an area of \SI{6e-15}{\meter\squared}, 
    and $\Theta=0$ otherwise. The direction of $\bm{P}$ lies along the easy axis so that there is no excitation at zero temperature since $\bm{m}_i\parallel \bm{P}$. Finite temperature, however, induces thermal fluctuations in magnetic moments $\bm{m}_i$ and therefore the net spin torque is finite, and consequently magnons are pumped into the AFMI layer. 

    \subsection{Measurement of spin accumulation}
    The magnon spin current is measured at detectors by means of the inverse spin Hall effect. The inverse spin Hall voltage is proportional to the spin accumulation at the interface of the detectors and the AFMI layer, given by \cite{ArneSpinPumping, PhysRevB.102.020408},
    \begin{equation} \label{eq:spinAccum}
        \bm{\mu}({d}) := G_r^{\uparrow \downarrow}  \left< \sum_{i}\big[\bm{m}_i(t,d)\times \dot{\bm{m}}_i(t,d)\big]\right>,
    \end{equation}
    where $G_r^{\uparrow \downarrow}$ %
    is the real part of the spin mixing conductance \cite{ArneSpinMixingConductance}, $\left< \cdot \right>$ denotes an average over both time and the ensemble, and $d$ is the distance between the injector and detector. 
    The time average for each ensemble member starts after steady state is reached, and afterwards, the ensemble average is taken over all spin accumulation signals. In our setup geometry (\cref{fig:OurSetup}), the inverse spin Hall detector measures only the $x$ component of the spin accumulation at the interface, $\mu_x$.

    The total spin accumulation along the direction of the magnon propagation can also be expressed as $\mu_x = N S_x$, where $N$ is the number of magnons that decay exponentially over distance due to Gilbert damping. Furthermore, $S_x=\hbar p \epsilon$, with reduced Planck constant $\hbar$, is the effective spin angular momentum of a magnon mode that is proportional to the helicity or handedness ($p=\pm 1$) and ellipticity ($0 \le \epsilon \le 1$ ). For linearly polarized magnon eigenmodes the ellipticity is zero, and thus $S_x=0$, while for circularly polarized magnon eigenmodes $S_x=\pm1$.

\section{Results}

    First, we numerically find the magnon dispersion relation of the AFMI layer and analyze the impact of a finite homogeneous DM interaction and transverse magnetic field on the magnon dispersion. Second, we establish a magnonic beating theory that describes spin transport based on the found magnon dispersion relations. Finally, we show numerical evidence to support our proposed theory. 
    
    \subsection{Magnon spectra}\label{sec:disprel}%
        In this part, we compute the magnon spectra of easy-plane hematite numerically using a standard approach \cite{proposalStandardMicromagneticproblem} at zero temperature and the absence of any spin torque, $\bm{H}_i^\mathrm{th}=\bm{H}_i^\mathrm{SOT}=0$.

        Magnons in the entire magnetic Brillouin zone can be excited by a magnetic field pulse with a spatial and temporal sinc function profile $\bm{h}(\bm{r},t)= \bm{h}_0 \text{sinc}(\bm{r})\text{sinc}(t)$, where $\bm{h}_0$ is the magnetic field vector. 
        Depending on the relative direction of the magnetic field pulse and magnetic moments, we can excite either the in-plane linearly polarized magnon modes, when $\mathbf{h}_0\|\hat{z}$, or out-of-plane linearly polarized magnon modes, when $\mathbf{h}_0\|\hat{y}$, in our sample geometry (see Fig. \ref{fig:OurSetup}). The in-plane mode is associated with the easy-axis anisotropy and has a lower magnon band gap while the out-of-plane mode is associated with the hard-axis anisotropy and has a higher magnon band gap in the easy-plane phase of hematite [see Eqs. (\ref{Eq:analyticalGaps1}) and (\ref{Eq:analyticalGaps2})].

\begin{figure}[H]
            \centering
            \begin{overpic}[width=0.23\textwidth]
            {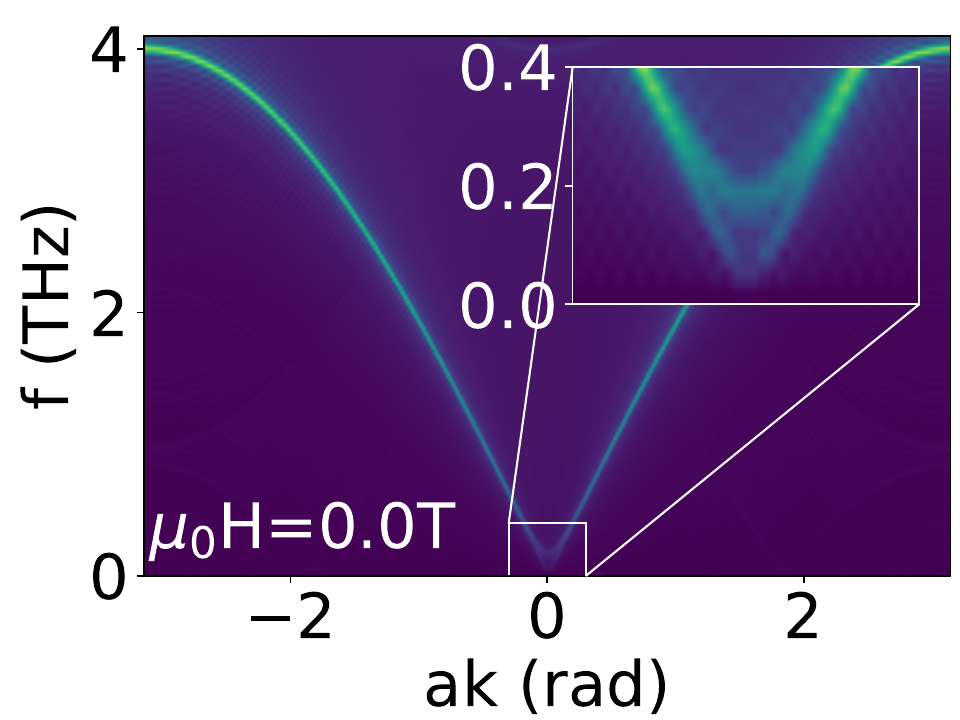}
            \put (90,20) {\textcolor{white}{(a)}}
            \put (15,75) {Without DM interaction}
            \end{overpic}
            \begin{overpic}[width=0.23\textwidth]{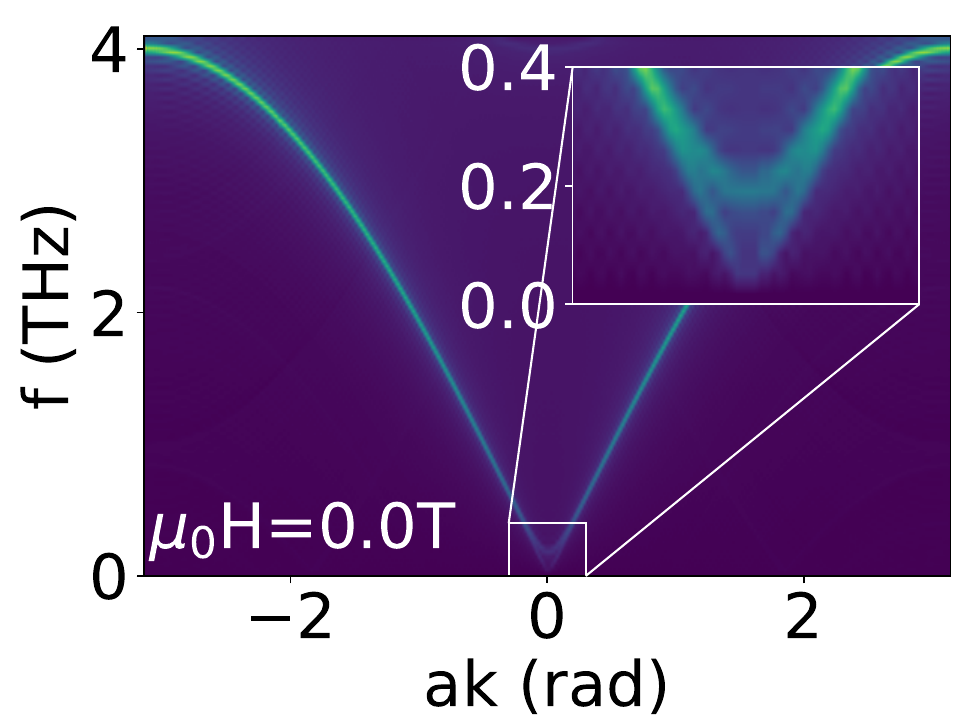}
            \put (90,20) {\textcolor{white}{(f)}}
            \put (20,75) {With DM interaction}
            \end{overpic}
            \begin{overpic}[width=0.23\textwidth]{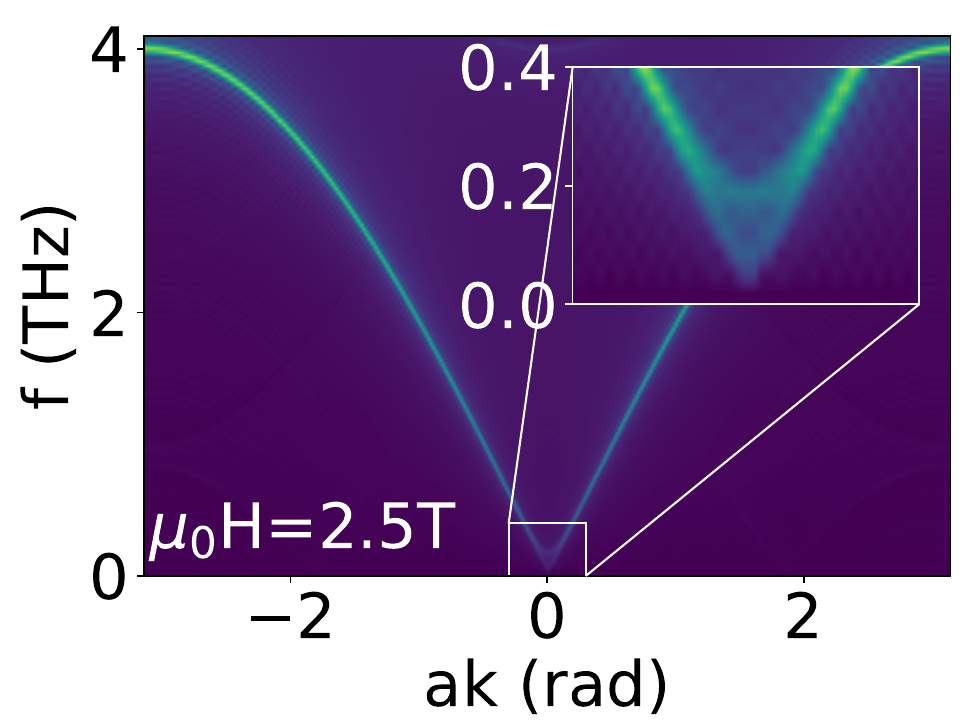}
            \put (90,20) {\textcolor{white}{(b)}}
            \end{overpic}
            \begin{overpic}[width=0.23\textwidth]{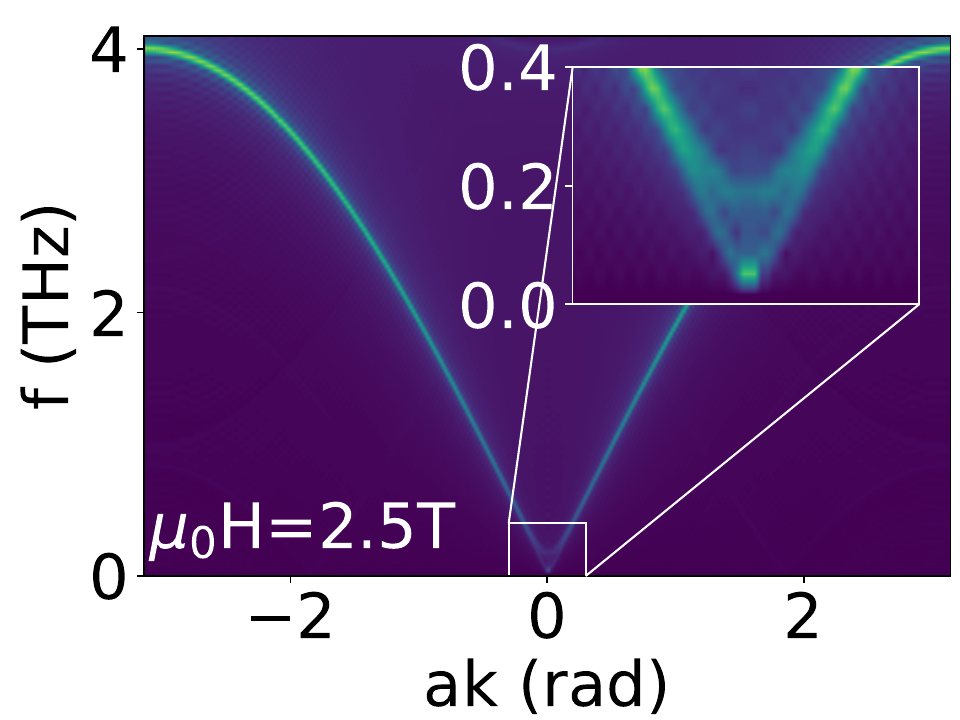}
            \put (90,20) {\textcolor{white}{(g)}}
            \end{overpic}
            \begin{overpic}[width=0.23\textwidth]{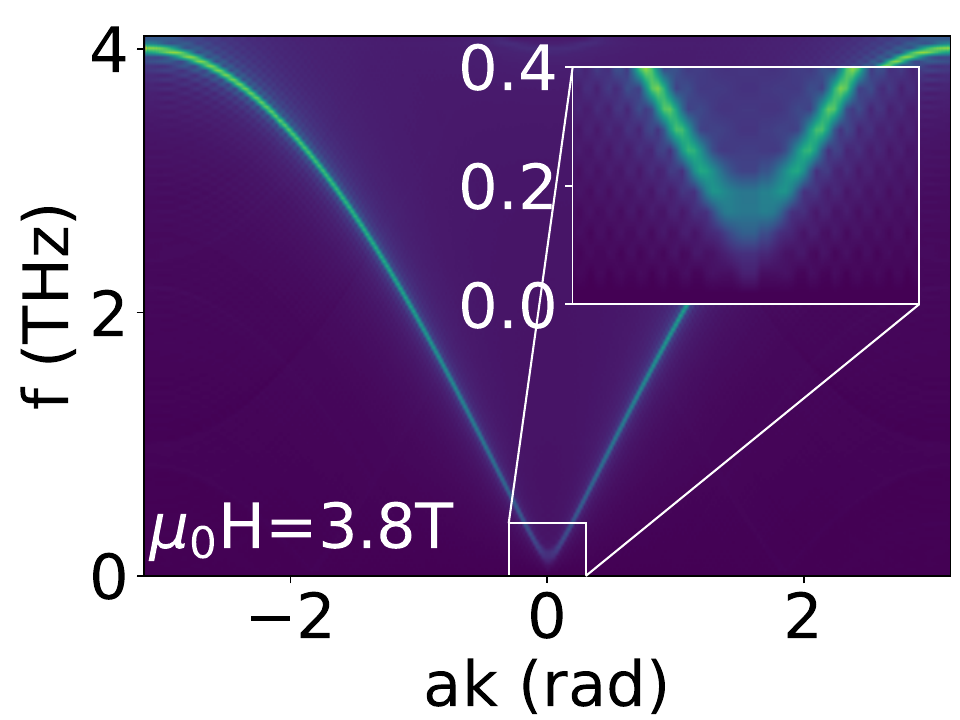}
            \put (90,20) {\textcolor{white}{(c)}}
            \end{overpic}
            \begin{overpic}[width=0.23\textwidth]{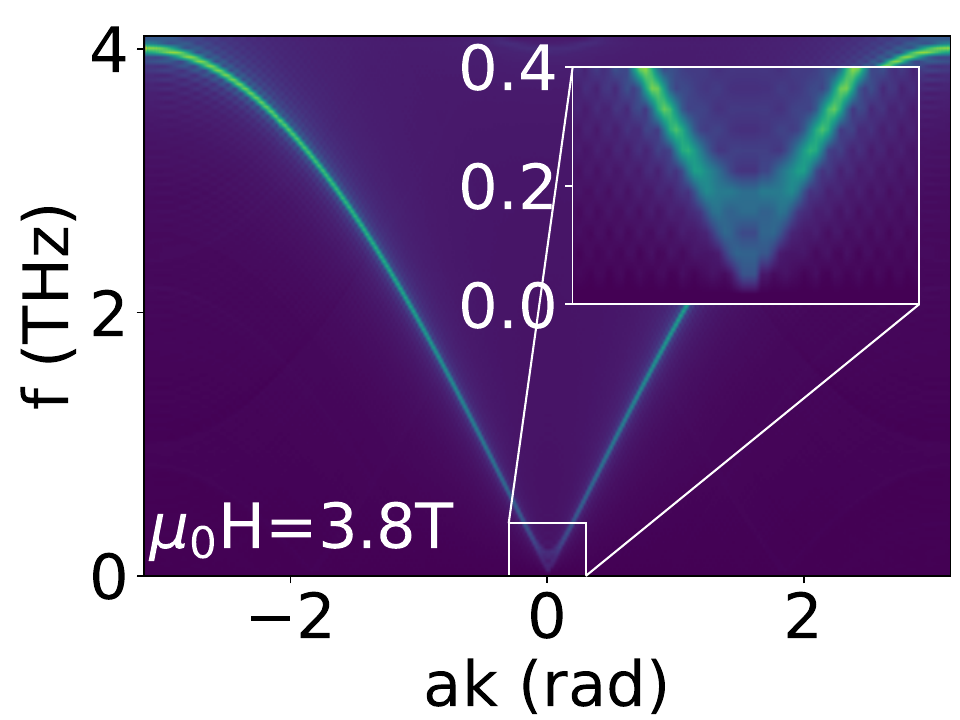}
            \put (90,20) {\textcolor{white}{(h)}}
            \end{overpic}
            \begin{overpic}[width=0.23\textwidth]{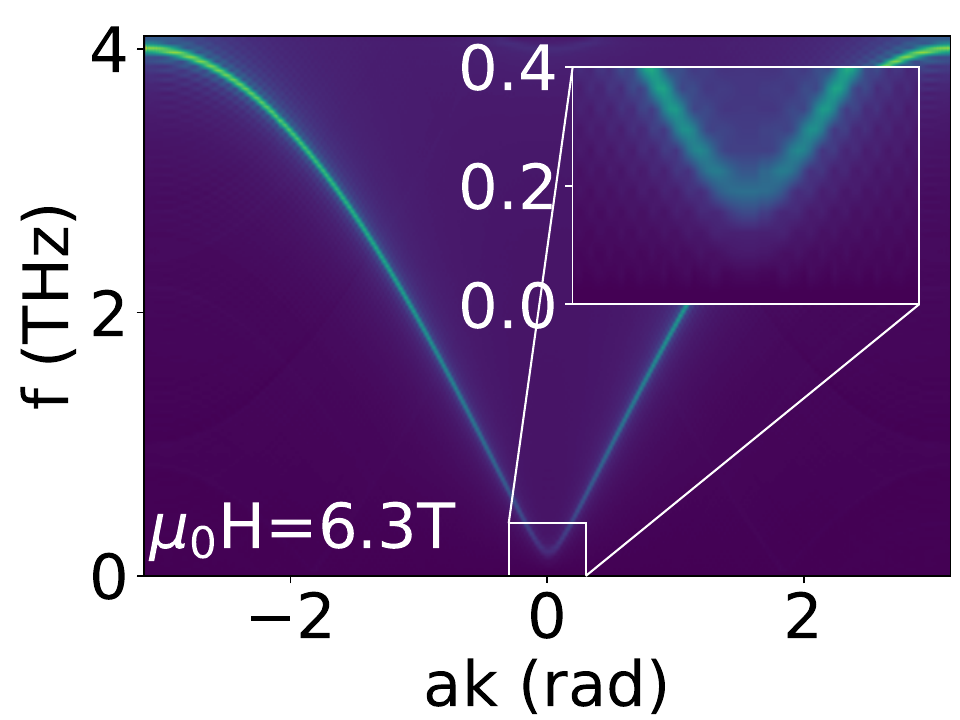}
            \put (90,20) {\textcolor{white}{(d)}}
            \end{overpic}
            \begin{overpic}[width=0.23\textwidth]{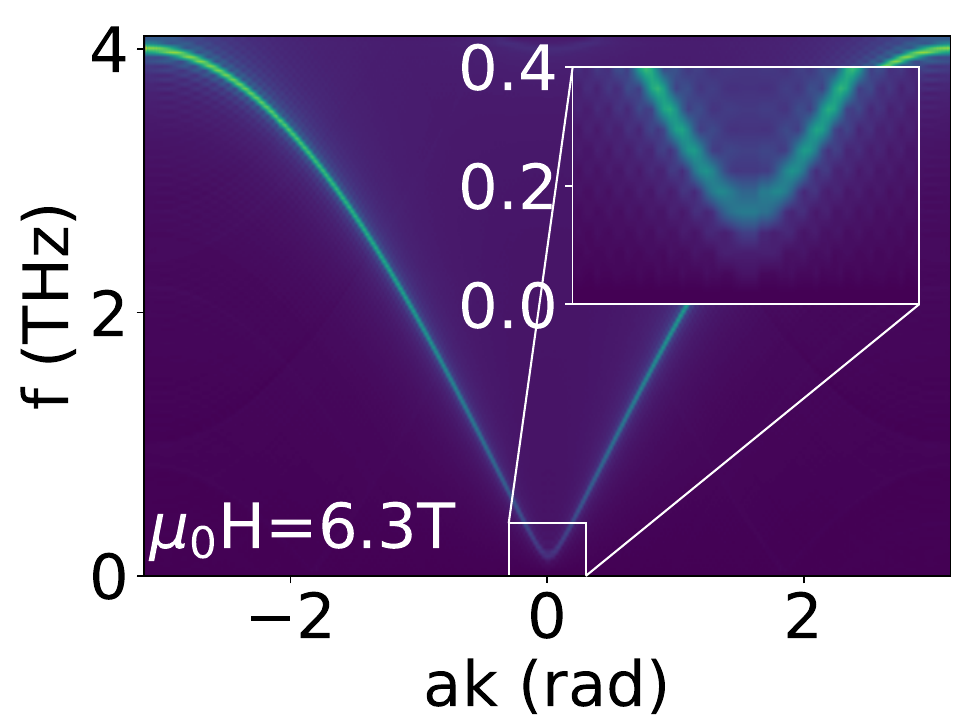}
            \put (90,20) {\textcolor{white}{(i)}}
            \end{overpic}
            \begin{overpic}[width=0.23\textwidth]{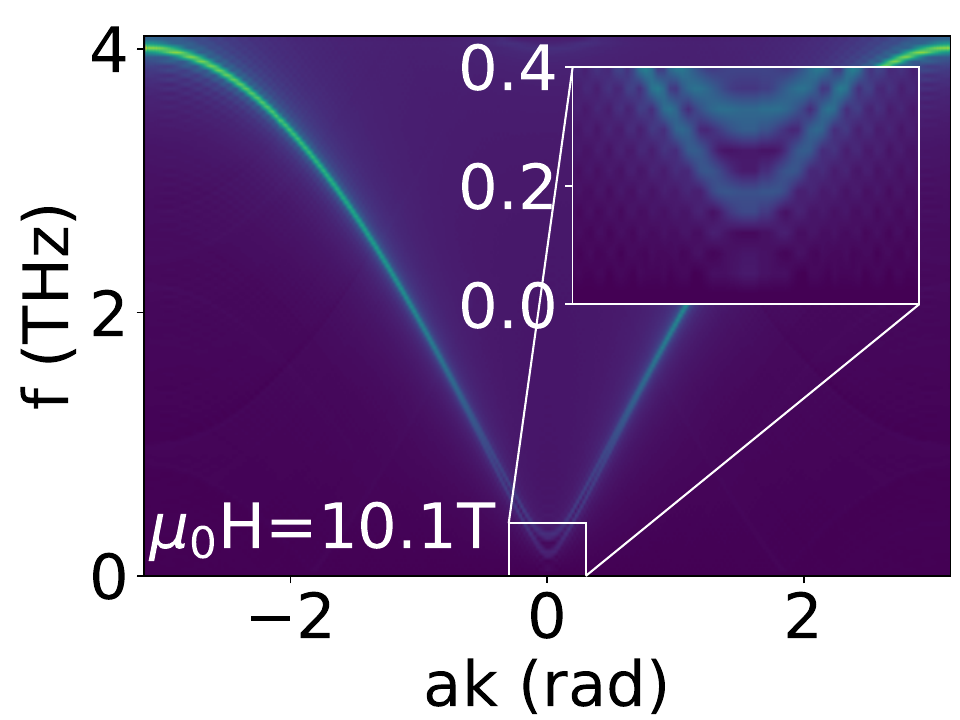}
            \put (90,20) {\textcolor{white}{(e)}}
            \end{overpic}
            \begin{overpic}[width=0.23\textwidth]{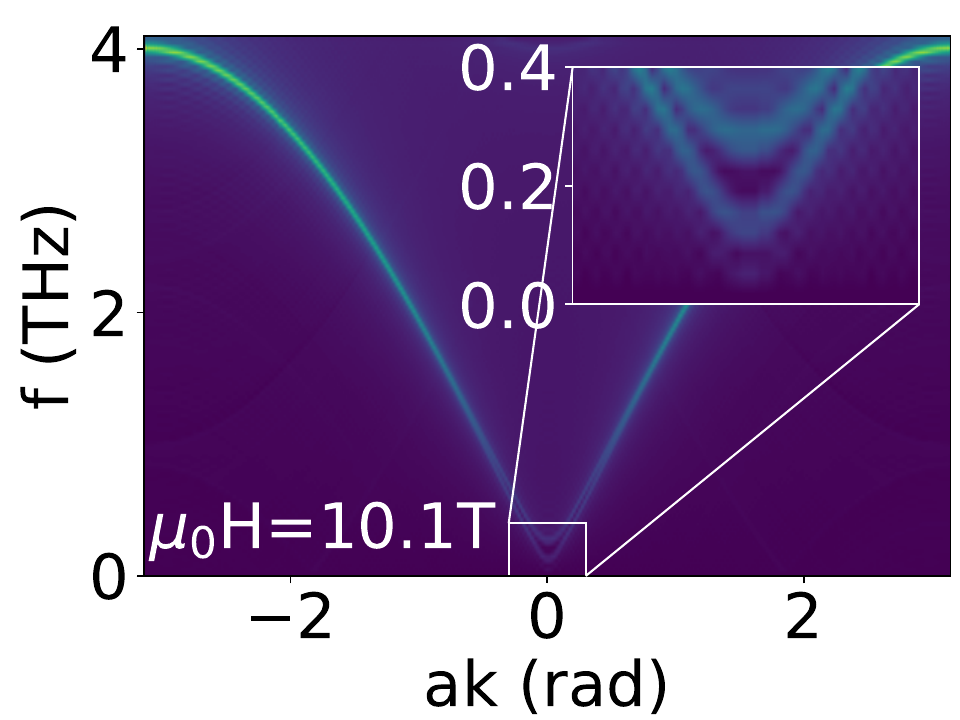}
            \put (90,20) {\textcolor{white}{(j)}}
            \end{overpic}
            \caption{Magnon dispersion relations for various transverse magnetic fields $H_y$ without a DM interaction (left column) and with a DM interaction (right column) at $T=0$. Without any transverse magnetic field [(a) and (f)], the two linearly polarized branches are separated by a band gap at the center of the Brillouin zone (see the inset). As the magnetic field increases, the bandgap of the lower branch increases [(b),(c) and (g),(h)], until at a critical field $H_y^c$ the bandgap between the two branches is closed and they become degenerate [(d) and (i)]. Within our material parameters for hematite, the critical field is around \SI{6}{T}. Above $H_y^c$, the order of branches is changed and a band gap reopens between the two branches [(e) and (j)]. }
            \label{fig:DispWithoutDMI}
        \end{figure}

\begin{figure}[h]
        \centering
        \includegraphics[width=0.8\linewidth]{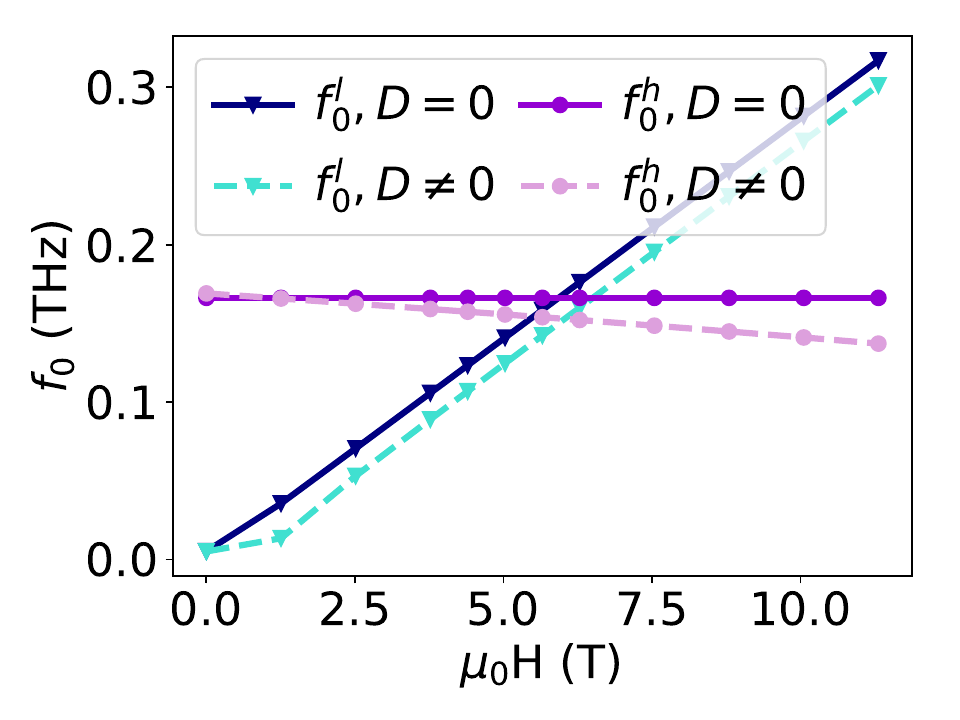}
        \caption{AFMR amplitude as a function of the transverse magnetic field  in the absence (solid lines) and presence (dashed lines) of a DM interaction. The intersection of the higher magnon branch $f^h$ in purple dots and the lower magnon branch $f^l$ in blue triangles shows the critical transverse magnetic field where the two magnon branches become degenerate. We conclude that the effect of a finite homogeneous DM interaction is negligible.}
        \label{fig:analyticGapsOfH}
        \end{figure}

        In \cref{fig:DispWithoutDMI}, we plot these two magnon modes related to two magnon polarization modes. Two magnon modes are almost degenerate at large wave vectors $k$, near the edge of magnetic Brillouin zone, while show different frequency values close to the center of magnetic Brillouin zone  ($k\approx 0$). The difference is pronounced at zero external magnetic field [see \cref{fig:DispWithoutDMI}(a) and (f)], where the magnon bandgap of the high-frequency branch (at around \SI{0.17}{THz}) is two orders of magnitude larger than the gap in the low-frequency branch (around \SI{5}{GHz}). However, the transverse magnetic field can tune the magnon bandgap and increase the lower magnon branch, while the high-frequency branch remains practically unchanged. Once the transverse magnetic field reaches some critical value $H_y^c$, both branches become degenerate in the entire Brillouin zone. At higher transverse magnetic fields, $H_y>H_y^c$, the order of the two branches swaps, so that the magnons of the low-frequency branch (i.e., in-plane oscillations) have a larger frequency than the magnons of the high-frequency branch (i.e., out-of-plane oscillations). 
        Within our set of parameters, the critical field can be read from the numerical dispersion relations as around $\mu_0 H_y^c \approx \SI{6}{T}$. 
        To obtain more insight into the nature and exact value of the critical transverse magnetic field, we compare numerical results with the magnon spectra calculated from the standard linear spin-wave theory  \cite{AlirezaContinuumModel,RezendeAFMReview}. The  magnon bandgaps or AFM resonance (AFMR) frequencies at $k=0$ for lower ($\text{l}$) and higher ($\text{h}$) magnon modes read
        \small\begin{align} 
                        f^{\text{l}}_0 &= \frac{\gamma_e}{2\pi M_s}\sqrt{16|a_\text{AFM}| K_\text{easy} + \left(M_s \mu_0 H_y \right)^2 \label{Eq:analyticalGaps1}- \frac{D}{2}M_s \mu_0 H_y},\\ 
                        f^{\text{h}}_0&= \frac{\gamma_e}{2\pi M_s}\sqrt{ 16|a_\text{AFM}|(K_\text{hard}+K_\text{easy}) \label{Eq:analyticalGaps2}  -\frac{D^2}{4} - \frac{D}{2} M_s \mu_0 H_y}.
        \end{align}
        \normalsize In \cref{fig:analyticGapsOfH}, we plot and compare the bandgaps of both magnon branches with and without the homogeneous DM interaction. 
        From \cref{Eq:analyticalGaps1} and \cref{Eq:analyticalGaps2}, we find the critical magnetic field, in which two magnon bands become degenerate, as $\mu_0 H^c_y = M_s^{-1}\sqrt{16|a_\text{AFM}| K_\text{hard}+{D^2}/{4}}$.
        Within the chosen material parameters, the critical magnetic field in the absence and the presence of a DM interaction is given by   $\mu_0 H^c_y(D=0) \approx \SI{5.9}{T}$ and $\mu_0 H^c_y($D=\SI{4.6}{\kilo\joule \per \meter\cubed}$) \approx \SI{6.0}{T}$, respectively. The critical field can also be read in \cref{fig:analyticGapsOfH} as the intersection.    
        We conclude that the effect of a homogeneous DM interaction on the magnon dispersion is negligible.
        Therefore, we will only consider the system without a homogeneous DM interaction in the rest of this paper.

        In summary, in this section, we have shown that the two magnon branches of an easy-plane AFMI can be modulated by a transverse magnetic field.

    \subsection{Coherent beating oscillation mechanism} \label{sec:beating}
The two magnon eigenmodes of the easy-plane hematite, shown in Fig. \ref{fig:DispWithoutDMI}, are linearly polarized and thus cannot carry any net spin angular momentum. However, we argue that in easy-plane AFMIs still a net spin angular momentum can be carried as shown experimentally.
A finite spin angular momentum can be carried by pairs of linearly polarized and orthogonal magnon eigenmodes with the same frequency but different wave numbers $k_1$ and $k_2$, excited on two magnon branches. Due to this finite wave-number difference, a spatial-dependent oscillation of the spin transport signal emerges that, analogously to the optical counterpart phenomenon, we call \textit{magnonic coherent beating oscillation}. The  difference between two magnon wave numbers with the same frequency can be controlled via an applied transverse magnetic field, as shown in the previous section, and thus the beating length can be changed. 
    
        The superposition of two linearly polarized and orthogonal magnon eigenmodes with wavenumbers $k_1$ and $k_2$ at the same frequency $f$, propagating along the $x$ direction, is given by
        \small\begin{align} \label{eq:beating}
             \Psi=& \frac{1}{\sqrt{2}}\left[\chi_Z \exp\left( i f t - i k_1 x  \right) + \chi_Y  \exp(i f t - i k_2 x \pm i \frac{\pi}{2})\right] \nonumber
              \\= &\frac{1}{\sqrt{2}}\chi_{ZY} \exp(i ft-i k_0x),
        \end{align}
        \normalsize 
        where 
        $\chi^T_Z=\left(
          \begin{array}{cccc}
            1 & 0 & 0 \\
          \end{array}\right)$ and 
        $\chi^T_Y=\left(
          \begin{array}{cccc}
            0 & 1 & 0 \\
          \end{array}\right)$
        are the transpose of the polarization eigenvectors for two linear polarization along $z$ (out-of-plane) and $y$ (in-plane) directions, respectively, while 
        $\chi^T_{ZY}=\left(
          \begin{array}{cccc}
            1 & e^{i \phi_k(x)} & 0 \\
          \end{array}\right)$
          is the superposition of them in the $ZY$ plane. We define $k_0=k_1+k_2$, $\Delta_k=k_1-k_2$, and $\phi_k(x) = \Delta_k x \pm \pi/2$. The effective spin angular momentum of the wave vector is given by 
          \begin{align}\label{eq:oscillatingSignal}
              S_x =\hbar p \epsilon= \hbar\expval{\hat{J}_x}{\Psi}=\hbar \sin\phi_k(x)
          \end{align} 
          where $\hat{J}_x = \hbar \tiny \begin{pmatrix} 0 & -i & 0 \\ i & 0 & 0 \\ 0 & 0 & 0 \end{pmatrix}\normalsize$  
          is the component of the spin-1 operator along the quantization axis in our geometry, i.e., the magnon transport direction. Therefore, the net spin angular momentum of the traveling magnon modes can  continuously vary between $S_z=+(-)\hbar$, for a right- (left-)handed circularly polarized wave, and $S_z=0$, for a linearly polarized wave, depending on $\phi_k(x)$. The distance $x_0$ at which the spin polarization sign is changed from left to right handed we call the magnonic beating length, $x_0 = \pi/\Delta_k$. This corresponds to the distance between a maximum and a minimum in the spin transport signal, with a zero transition at $x_0/2$, as shown in \cref{fig:distDepSpinAccumPerField}.

    \subsection{Band-resolved magnon population} \label{sec:numbeating}
        In the previous section, we have introduced a coherent beating mechanism that is based on the excitation of pairs of magnons, where each magnon belongs to one of the branches in the dispersion relation. In order to show the evidence of these magnon pairs numerically, we compute the magnon population at low, but finite, temperature and a finite spin torque $\bm{H}_i^\text{SOT}>\bm{0}$ in the injector region. By Fourier transforming the temporal and spatial-dependent spin configuration, we find the occupied magnon modes that contribute to the long-distance spin transport. 

\begin{figure}[htbp]
            \centering
            \begin{overpic}[width=0.23\textwidth]{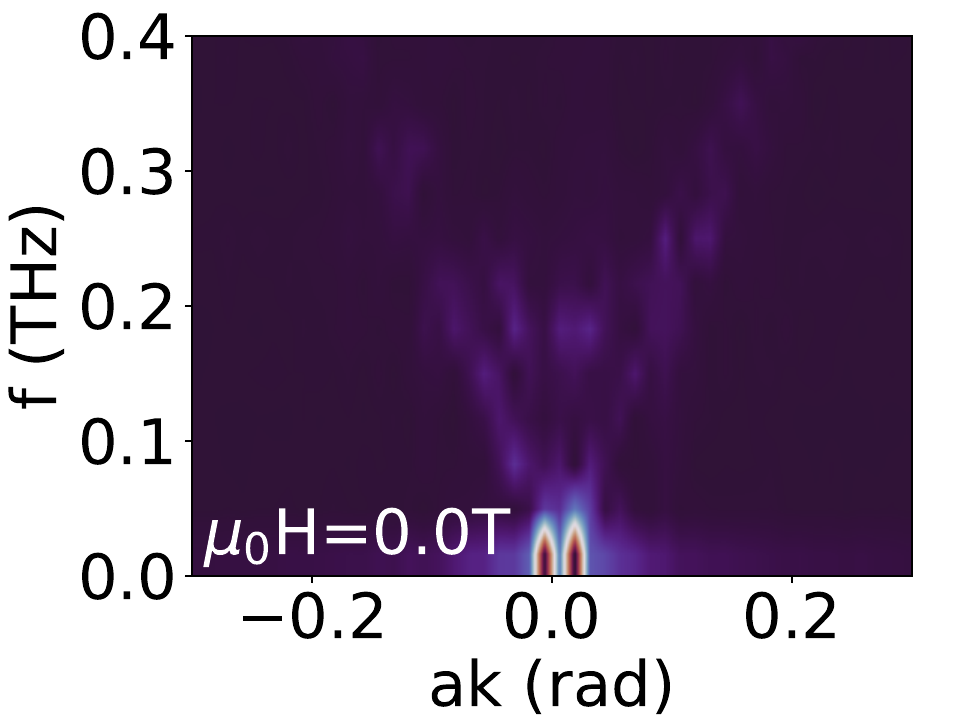}
            \put (85,20) {\textcolor{white}{(a)}}
            \end{overpic}
            \begin{overpic}[width=0.23\textwidth]{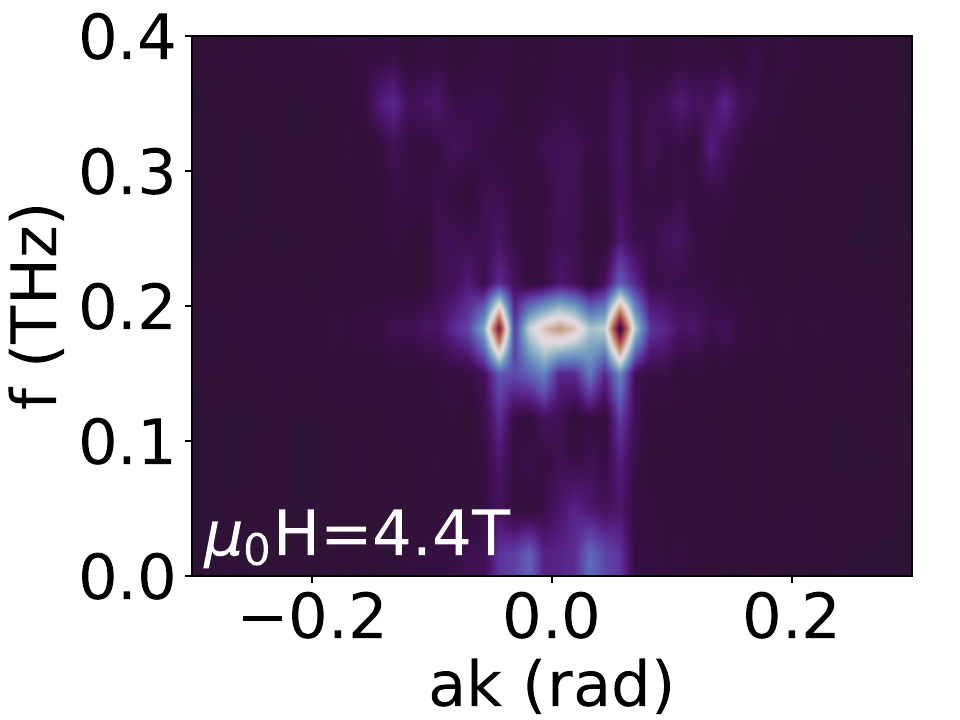}
            \put (85,20) {\textcolor{white}{(d)}}
            \end{overpic}
            \begin{overpic}[width=0.23\textwidth]{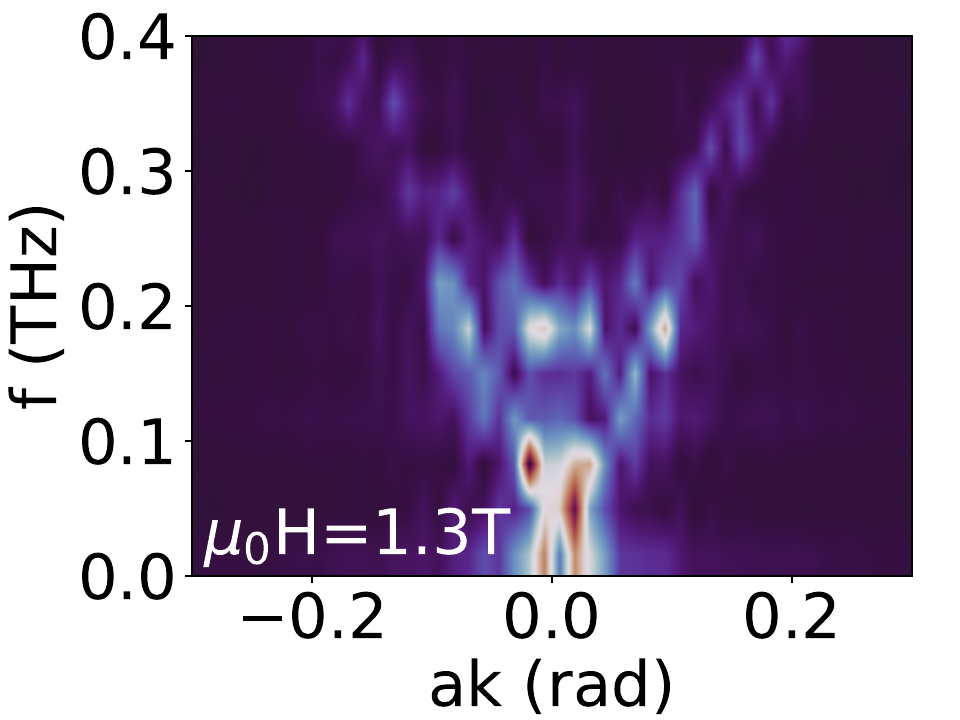}
            \put (85,20) {\textcolor{white}{(b)}}
            \end{overpic}
            \begin{overpic}[width=0.23\textwidth]{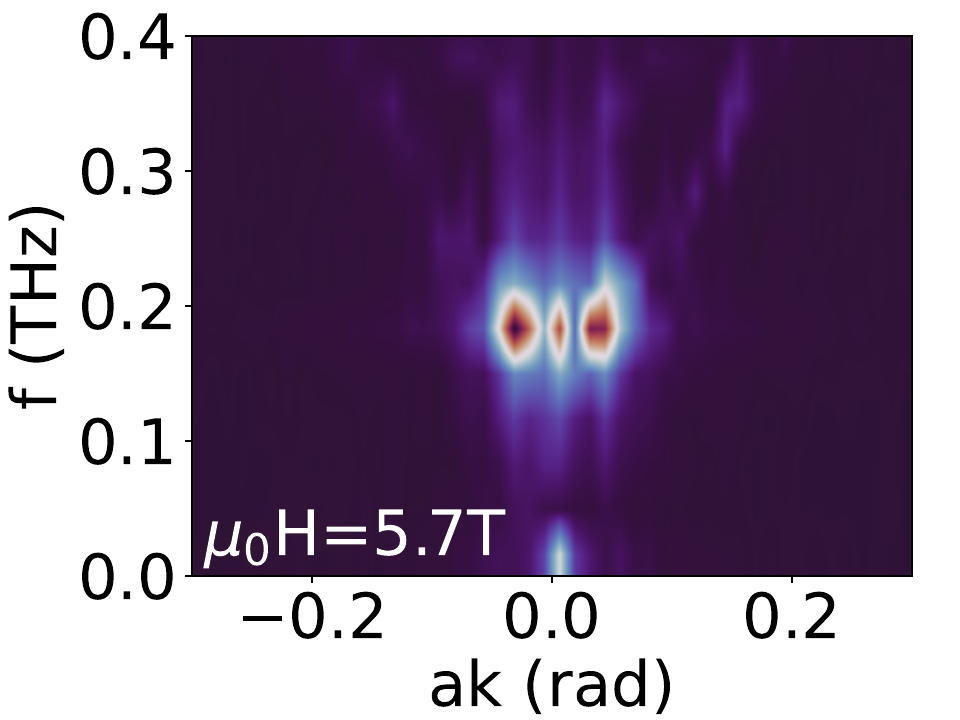}
            \put (85,20) {\textcolor{white}{(e)}}
            \end{overpic}
            \begin{overpic}[width=0.23\textwidth]{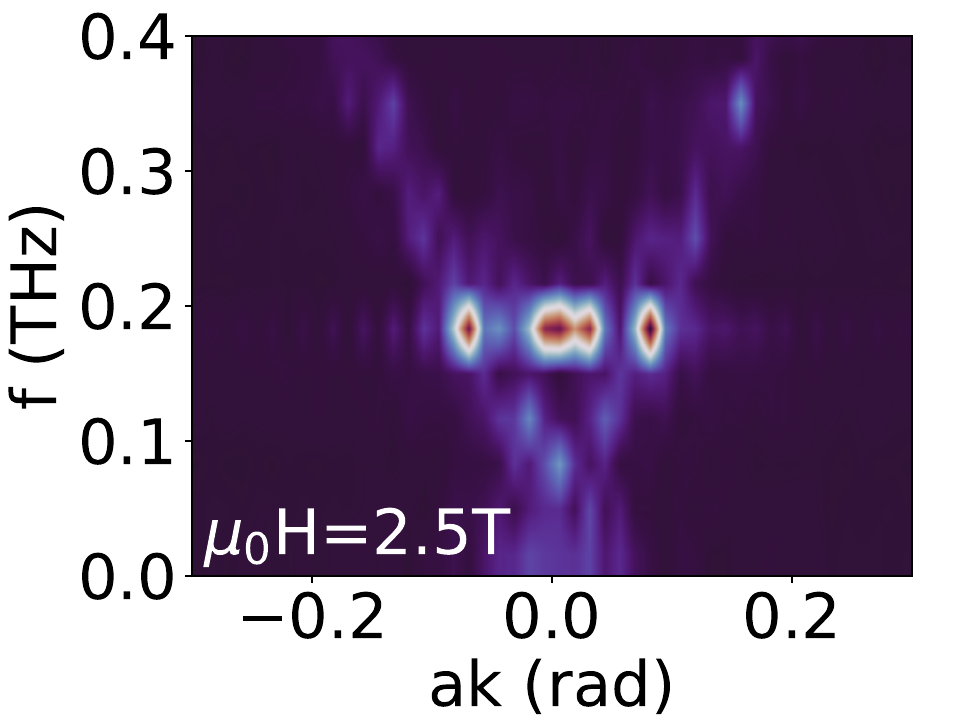}
            \put (85,20) {\textcolor{white}{(c)}}
            \end{overpic}
            \begin{overpic}[width=0.23\textwidth]{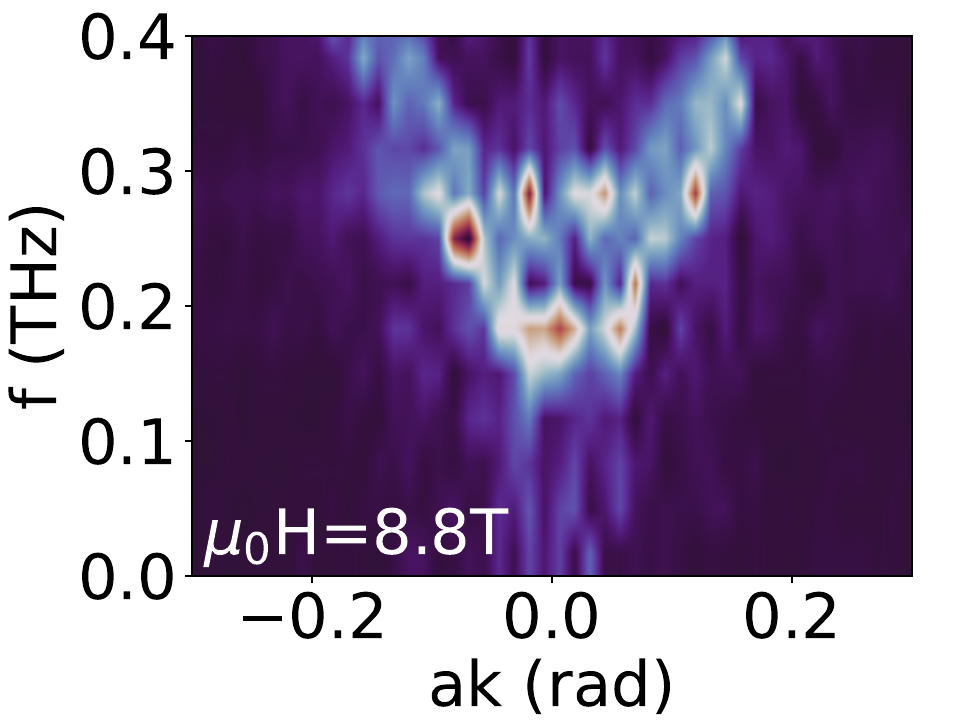}
            \put (85,20) {\textcolor{white}{(f)}}
            \end{overpic}
            \caption{Magnon modes contributing to the spin transport signal for various transverse magnetic fields at finite temperature and in the presence of the spin Hall torque. Two pairs of occupied magnon modes show up around the magnetic Brillouin zone center at intermediate strengths of the transverse magnetic field [(c) to (e)]. Each pair consists of two modes, one on each branch, at the same frequency \SI{0.18}{THz}, but different wave numbers. The color map refers to the signal strength, which is proportional to the magnon occupation number, from dark meaning very low over white to red meaning very high intensity.}
            \label{fig:DispInsetFew}
        \end{figure}

        For our choice of parameters, i.e., the amplitude of spin torque and temperature, only low-energy magnons are excited. Thus, only the center of the Brillouin zone is shown in \cref{fig:DispInsetFew}.

        For small transverse magnetic fields [see \cref{fig:DispInsetFew}(a) and (b)], and above the critical magnetic field [see \cref{fig:DispInsetFew}(f)] mostly the modes in the lower magnon branch are occupied. These are linearly polarized magnon modes that cannot carry any net spin angular momentum. At these transverse magnetic field strengths, we observe a low and rapidly decaying spin transport signal (not shown). 

        At intermediate transverse magnetic field strengths, however, pairs of magnons with the same frequency that belong to two different magnon branches appear [see \cref{fig:DispInsetFew}(c)-(e)]. The difference in the wave numbers between two branches at one frequency, $\Delta_k=k_1-k_2$, becomes smaller as the transverse magnetic field reaches the critical field due to the band modulation shown in \cref{sec:disprel}.

    \subsection{Distance-dependent spin accumulation signal}\label{sec:transport}
        Finally, we show the spin transport data and connect it to the proposed coherent beating oscillation theory. As we have already discussed, the detected spin signal can be written as $\mu_x(d) = \hbar N p \epsilon = \hbar N(d) \sin\phi_k(d)$. The amplitude of the spin signal is proportional to the number of magnons $N(d)$ that decay exponentially with the distance $d$ because of the Gilbert damping, and is modulated by the magnonic beating parameter $\sin\phi_k(d)$ with $\phi_k(d)= \Delta_k d \pm \pi/2$.
        
\begin{figure}
            \centering
            \includegraphics[width=1.0\linewidth]{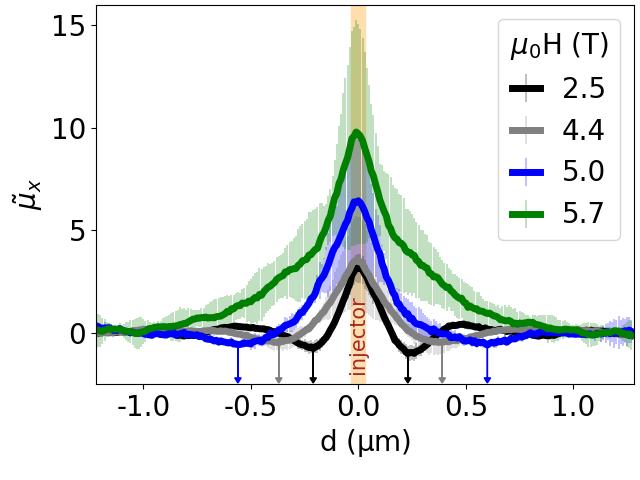}
            \caption{Dimensionless spin accumulation signal $\tilde{\mu}_x=\mu_x(H_y,d)/\mu_x(H_y=0,d=0)$ along the easy-axis direction as a function of the distance between detectors and the injector, placed at the center. The transverse magnetic field $H_y$ modulates the shape of the spin transport signal. At low magnetic fields, the sign of the spin signal polarization changes at intermediate distances. At the critical transverse magnetic field $H_y^c$ (green curve), where the two magnon branches are degenerate, the spin signal is always positive and decays exponentially.}
            \label{fig:distDepSpinAccumPerField}
        \end{figure}

        Figure \ref{fig:distDepSpinAccumPerField} displays the numerically found distance dependent spin accumulation in the nonlocal geometry for four different values of the transverse magnetic field $H_y$. Each data set is computed with an ensemble average over 20 realizations, and the uncertainty environment corresponds to the standard deviation.  
        
        As expected from the geometry, we have $\sin\phi_k(d=0) = 1$ and thus the detected spin signal is the highest close to the center, where the injector pumps spin angular momentum into the system. However, the amplitude of the spin accumulation signal does not decay exponentially, as expected from the diffusive magnon transport theory, at some transverse magnetic fields. It rather shows a damped oscillating behavior, and changes the sign periodically. At these fields, the magnon bands are non-degenerate, and thus a finite $\Delta_k$ leads to a finite beating length $x_0$, which increases with increasing magnetic field (since $\Delta_k$ decreases). In order to give a quantitative example, at $\mu_0 H_y=\SI{2.5}{T}$, we have $\Delta_k\text{a} = 0.06$ from \cref{fig:DispInsetFew}(c), and estimate $x_0$ to be $x_0(\SI{2.5}{T})=\SI{260}{nm}$, which is in line with the minimum of the black curve.
    
        On the other hand, at the critical magnetic field strength of $\mu_0 H_y \approx\SI{5.7}{T}$, where the two magnon branches are degenerate and thus the beating length diverges, $\sin(\Delta_k \to 0) = 1$, the spin signal does not change the sign and only decays exponentially.

\section{Summary and Conclusion} \label{sec:conclusion}
    We have found a long-distance and tunable spin transport signal in orthorhombic easy-plane AFMIs using micromagnetic simulations, in agreement with recent experimental measurements. Our model represents a large class of easy-plane AFM materials, including hematite above the Morin transition.
    
    We demonstrated how a \textit{finite spin  signal} and its \textit{helicity} and \textit{amplitude} can be modulated by a transverse magnetic field using a \textit{coherent beating oscillation} between two linearly polarized magnon eigenmodes. 
    Based on our theoretical framework and numerical experiments, we argue that this behavior is a generic feature of all easy-plane AFMIs with two orthogonal linearly polarized eigenmodes and not only weak ferromagnets or canted AFMIs with finite homogeneous DM interaction, such as hematite.
    
\section*{Acknowledgment}
This project has been supported by the Norwegian Financial Mechanism Project No. 2019/34/H/ST3/00515, ``2Dtronics''; and partially by the Research Council of Norway through its Centres of Excellence funding scheme, Project No. 262633, ``QuSpin''.
The team in Mainz acknowledges support by the DFG (SFB SPIN+X No. 268565370).

\bibliography{literature}


\end{document}